%% file: main.tex
\documentclass[sigconf]{acmart}
\usepackage{multirow}
\usepackage{tablefootnote}
\usepackage{enumitem}
\usepackage[font=scriptsize]{caption}

\AtBeginDocument{%
  \providecommand\BibTeX{{%
    \normalfont B\kern-0.5em{\scshape i\kern-0.25em b}\kern-0.8em\TeX}}}

\copyrightyear{2020}
\acmYear{2020}
\setcopyright{acmcopyright}\acmConference[CIKM '20]{Proceedings of the 29th ACM International Conference on Information and Knowledge Management}{October 19--23, 2020}{Virtual Event, Ireland}
\acmBooktitle{Proceedings of the 29th ACM International Conference on Information and Knowledge Management (CIKM '20), October 19--23, 2020, Virtual Event, Ireland}
\acmPrice{15.00}
\acmDOI{10.1145/3340531.3412754}
\acmISBN{978-1-4503-6859-9/20/10}

\begin{document}
\fancyhead{}

\title{GraphSAIL: Graph Structure Aware Incremental Learning for Recommender Systems}



\author{Yishi Xu}
\authornote{Work done as interns at Huawei Noah's Ark Lab Montreal Research Center.}
\email{yishi.xu@umontreal.ca}
\affiliation{%
  \institution{Mila, Universit\'{e} de Montr\'{e}al}
  \city{Montreal}
  \state{QC}
  \country{Canada}
} 

\author{Yingxue Zhang}
\email{yingxue.zhang@huawei.com}
\affiliation{%
  \institution{Huawei Noah's Ark Lab\\Montreal Research Center}
  \city{Montreal}
  \state{QC}
  \country{Canada}
}

\author{Wei Guo}
\email{guowei67@huawei.com}
\affiliation{%
  \institution{Huawei Noah's Ark Lab}
  \city{Shenzhen}
  \country{China}
}

\author{Huifeng Guo}
\email{huifeng.guo@huawei.com}
\affiliation{%
  \institution{Huawei Noah's Ark Lab}
  \city{Shenzhen}
  \country{China}
}

\author{Ruiming Tang}
\email{tangruiming@huawei.com}
\affiliation{%
  \institution{Huawei Noah's Ark Lab}
  \city{Shenzhen}
  \country{China}
}

\author{Mark Coates}
\email{mark.coates@mcgill.ca}
\affiliation{%
  \institution{McGill University}
  \city{Montreal}
  \state{QC}
  \country{Canada}
}

\renewcommand{\shortauthors}{Yishi Xu, et al.}

\begin{abstract}
Given the convenience of collecting information through online services, recommender systems now consume large scale data and play a more important role in improving user experience. With the recent emergence of Graph Neural Networks (GNNs), GNN-based recommender models have shown the advantage of modeling the recommender system as a user-item bipartite graph to learn representations of users and items. However, such models are expensive to train and difficult to perform frequent updates to provide the most up-to-date recommendations. In this work, we propose to update GNN-based recommender models incrementally so that the computation time can be greatly reduced and models can be updated more frequently. We develop a Graph Structure Aware Incremental Learning framework, \textbf{GraphSAIL}, to address the commonly experienced catastrophic forgetting problem that occurs when training a model in an incremental fashion. Our approach preserves a user's long-term preference (or an item's long-term property) during incremental model updating. GraphSAIL implements a graph structure preservation strategy which explicitly preserves each node's local structure, global structure, and self-information, respectively. We argue that our incremental training framework is the first attempt tailored for GNN based recommender systems and demonstrate its improvement compared to other incremental learning techniques on two public datasets. We further verify the effectiveness of our framework on a large-scale industrial dataset.


\end{abstract}

\ccsdesc[500]{Information systems~Recommender systems}
\ccsdesc[500]{Computing methodologies~Online learning settings}
\keywords{incremental learning, catastrophic forgetting, recommendation system, graph neural networks, representation learning}

\maketitle

\input{sections/introduction.tex}

\input{sections/related_work.tex}

\input{sections/preliminaries.tex}

\input{sections/methodology.tex}
\input{sections/experimental_results.tex}
\input{sections/application.tex}

\input{sections/conclusion.tex}

\bibliographystyle{ACM-Reference-Format}
\bibliography{main}

\end{document}

%% file: sections/introduction.tex
\section{Introduction}

The amount of online information has been growing explosively. To help users identify what might interest them from a gigantic and rapidly expanding pool of candidates, recommender systems play an important role in many information retrieval (IR) tasks, including web search, news recommendation, and online advertising. Nowadays, deep learning models attract attention in the recommendation community due to their superiority for feature representation in the computer vision and nature language processing fields. 
With careful design of the neural architectures, many industrial companies deploy deep recommendation models in their commercial systems, such as Wide \& Deep~\cite{cheng2016wide} in Google Play, DeepFM~\cite{deepfm} and PIN~\cite{pin} in Huawei AppGallery.


\begin{figure}
    \centering
    \includegraphics[width=0.65\linewidth]{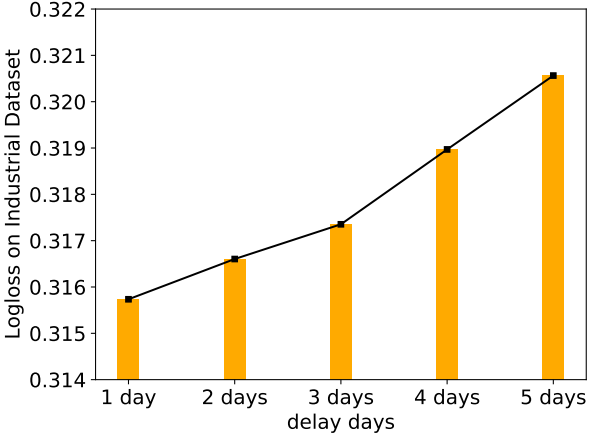}
    \caption{Model performance degrades when the model stops updating. x-axis: number of days with no update; y-axis: normalized logloss (smaller is better).}
    \label{fig:delay}
 \vspace{-0.5cm}
\end{figure}

To achieve good performance, deep recommendation models need to be trained on a huge volume of training data for multiple epochs until convergence. Industrial level recommendation systems collect users' interaction histories every day. It is critical to construct a model that can accurately capture both a user's long-term and short-term preference. In many practical settings, a fixed-size time window of training data (e.g. recent 30 days) is used to train deep recommendation models. The large volume of training data results in a long training time, meaning that it is impossible to include the most recent data in the training. The most recent data play a crucial role when a model tries to capture a user's dynamic preferences in order to achieve satisfying recommendation performance. We observe performance degradation when the recommendation model stops updating in a mainstream App Store, as presented in Figure~\ref{fig:delay}. For instance, if the model stops updating for 5 days, the model performance degrades 1.53\% in terms of normalized logloss, which may lead to significant loss of revenue and a poor user experience. However, directly retraining the model using only the recent records often causes a catastrophic forgetting problem, with the model losing track of the key user information needed to capture long term preferences.


Incremental learning provides one direction for tackling this problem. Incremental learning~\cite{kirkpatrick2017overcoming_forgetting, Shmelkov_2017_ICCV, End-To-End,rebuffi2017icarl, mallya2018packnet_IL, xu2018reinforced_IL, gag_streaming_gnn} uses the most recent data to update the current model, but is designed to prevent substantial forgetting. This significantly improves training efficiency without extra computational resource and meanwhile prevents model performance degradation. There are three main lines of work for incremental learning: \textit{experience replay} (reservoir), \textit{parameter isolation} and \textit{regularization based} methods. In this work, we mainly investigate the regularization based approach.

Recently, with the developments and improvements of graph neural networks (GNNs)~\cite{kipf2017,gilmer2017,zhang2019_bgcn}, graphs have proven to be among the best performing mechanisms for modeling the abundant relationship information present in recommendation datasets. User-item interaction graphs~\cite{ying2018,gcmc_vdberg2018,NGCF_wang19} and item-item co-occurrence graphs~\cite{graph-din} have been employed and shown to alleviate the data sparsity and cold start problems and improve recommendation relevance~\cite{ying2018,NGCF_wang19,graph-din,sun2020_mgcf}.  However, as is the case for deep recommendation models, graph neural network (GNN) based recommendation systems usually suffer from a low training efficiency. To the best of our knowledge, there is no existing incremental learning technique that is specifically tailored for GNN based recommendation system models. The design of an effective system is thus a key step towards allowing such models to make use of the most recent data. 

Incremental learning in recommendation models faces distinct challenges, because the data are of a multi-field categorical form. Furthermore, when distilling knowledge in graph-structured data we must be careful to preserve graph
structure information.
To address the above challenges, in this work, we provide a novel incremental training paradigm for graph neural network based recommender systems using knowledge distillation~\cite{hinton2015distilling}. Our distillation can be categorized into three components: (1) \textit{local structure distillation}; (2) \textit{global structure distillation}; (3) \textit{self-embedding distillation}. Self-embedding distillation prevents drastic changes of each individual embedding vector. To preserve the topological information, we propose local and global structure distillation modules that permit knowledge transfer that explicitly accounts for the topological semantics.

To summarize, this work makes the following three main contributions:
\begin{enumerate}[label=(\arabic{enumi}),ref=\arabic{enumi}, wide=0pt, topsep=4pt]
    \item GraphSAIL is the first attempt to perform incremental learning on GNNs for recommender systems. The proposed method achieves competitive performance and computation efficiency while at the same time alleviating the catastrophic forgetting issue.    
    \item We propose a novel strategy for preserving both local structure and global structure in GNN models. We achieve local structure preservation by encouraging embedding similarity between corresponding local graph neighborhoods in successively learned models. We use probability distributions to represent the global structure information, and minimizing the distance between these distributions for successive models enables topology-aware knowledge transfer.     
    \item The proposed solution has excellent generalization ability and is readily extendable to any GNN-based recommender system.
\end{enumerate}

%% file: sections/related_work.tex
\section{Related work}
\subsection{Graph-based Recommendation}
In the collaborative filtering and recommender system literature, there has been a considerable research effort focusing on developing techniques to employ graph representations. Early works~\cite{item-rank,hop-rec} used label propagation and random walks on the user-item interaction graph to derive similarity scores for user-item pairs. Recent works started to apply GNNs~\cite{ying2018,NGCF_wang19,sun2020_mgcf,DBLP:conf/aaai/MaMZSLC20}. PinSage~\cite{ying2018} uses a simple GNN (originally called GraphSage~\cite{hamilton2017}) on the item-item graph and reportedly provides significant performance gains for the Pinterest recommendation system. NGCF~\cite{NGCF_wang19} utilizes the user-item interaction graph and captures the relationship between a central node and its neighbours by aggregating neighborhood information. 
MGCCF~\cite{sun2020_mgcf} leverages GNNs to learn user-item, user-user and item-item graph structures simultaneously.
Another line of work deals with the complex and heterogeneous interaction types between user and item in large-scale e-commerce networks~\cite{ cen2019representation_multiplex, fan2019metapath_rec}. 
However, all such models assume that the entire dataset is available for batch training, while our work focuses on incrementally training models using new data only.

\subsection{Incremental Learning}
Despite the success of deep learning models, they suffer from catastrophic forgetting when new classes/tasks and data are added into the training pipeline incrementally~\cite{End-To-End,kirkpatrick2017overcoming_forgetting}. Various incremental learning techniques have been introduced to combat this issue. There are three main lines of work for incremental learning: experience replay(reservoir), parameter isolation and regularization based methods. 
Reservoir methods use an additional data reservoir to store the most representative historical data and replay it while learning new tasks to alleviate forgetting~\cite{rebuffi2017icarl}. 
Parameter isolation trains distinct models for different tasks, but leads to continual growth of the model size which is not favourable for training large-scale systems~\cite{mallya2018packnet_IL, xu2018reinforced_IL}. 
Regularization based approaches aim to consolidate previous knowledge by introducing regularization terms in the loss when learning on new data\cite{ Shmelkov_2017_ICCV, End-To-End, kirkpatrick2017overcoming_forgetting, yang2019adaptive}.

In the field of recommendation, several approaches have been proposed for incremental model training~\cite{yu2016incremental_big_data,li2019sparse_online_cf}. However, most of them either use matrix factorization as the base model or impose an $L_2$ penalty as the regularization technique. 
Other recent attempts that use knowledge distillation for GNN models~ \cite{zhou2020continual,yang_distillatingGCN} are not directly suitable for recommendation scenarios. In this work, we investigate practical regularization based incremental learning approaches for GNN-based recommendation models.

%% file: sections/preliminaries.tex
\section{Preliminaries}
Our model is a data efficient training framework built on top of existing GCN based recommendation systems such as~\cite{ying2018,sun2020_mgcf} and knowledge distillation~\cite{hinton2015distilling}. We briefly review these two components in this section.

\subsection{GCN-based Recommender Models}

In a recommendation scenario, the user-item interaction can be formulated as a bipartite graph with two types of nodes. We consider Pinsage~\cite{ying2018} as an example. Its learning process consists of two training phases:~\emph{forward sampling} and~\emph{backward aggregating}. 
After sampling nodes' neighbours from layers $1$ to $K$, the GCN model encodes the user and item nodes by iteratively aggregating $k$-hop neighborhood information via graph convolution. The aggregation step allows the model to learn rich semantic information for both user nodes and item nodes from multiple interaction records. There are initial embeddings $\mathbf{e}_u$ and $\mathbf{e}_v$ that are learned for each user $u$ and item $v$. These embedding are learned at the same time as the parameters of the GCNs. 
The layer-$k$ embedding of the target user $u$ can be represented as:
\begin{align}
\small
 \mathbf{h}^k_{u} &= \sigma\Big(\mathbf{W}_u^k\cdot \textbf{[}
 \mathbf{h}_u^{k-1}; \mathbf{h}_{\mathcal{N}(u)} ^{k-1}
 \textbf{]}\Big),
 \hspace{0.2cm}{\mathbf{h}}^0_{u} = {\mathbf{e}}_u \, ,\\
\mathbf{h}_{\mathcal{N}(u)} ^{k-1} &=  \mathrm{AGGREGATOR_u}\Big(
 \left\{\mathbf{h}^{k-1}_{v},  v \in \mathcal{N}(u)\right\} \Big) \,.
\end{align}
Here $\mathbf{e}_u$ is the initial user embedding, $[\,\,;\,]$ denotes concatenation, $\sigma(\cdot)$ is the
$\mathrm{ReLU}$ activation function, and $\mathbf{W}_u^k$ is the
layer-$k$ (user) transformation weight matrix shared across all user nodes. $\mathbf{h}_{\mathcal{N}(u)} ^{k-1}$ is the learned neighborhood embedding. $\mathrm{AGGREGATOR_u}$ is the aggregation function designed for the user nodes. Similar expressions apply for the item embedding. After aggregation at the $K$-th layer, the final user and item embedding are denoted as $\mathbf{emb}_{u}$ and $\mathbf{emb}_{i}$.

\subsection{Knowledge Distillation}

Knowledge distillation techniques were proposed initially to transfer knowledge from a large and complex model into a smaller distilled model~\cite{hinton2015distilling,yang_distillatingGCN}. Subsequently, it became common to use knowledge distillation to address the catastrophic forgetting issue in incremental training~\cite{Shmelkov_2017_ICCV, End-To-End}. The essence of knowledge distillation for incremental learning is to use a \textit{teacher model} trained on the data acquired from the old tasks or history data and a \textit{student model} trained on the knowledge acquired from the new tasks or new data. When training the student, a distillation metric is applied to the loss, in order to retain the knowledge acquired by the teacher model. In our work, in contrast to existing knowledge distillation methods that focus only on the prediction or the middle activation, GraphSAIL explicitly distills knowledge in order to preserve the graph topological information learned by the teacher model and transfer it to the student model.

%% file: sections/methodology.tex
\section{Methodology}
In this section, we introduce the three general components we propose for incrementally training a GNN-based recommendation model.  \emph{First}, a \textbf{local structure distillation} mechanism is proposed to preserve a user's long-term preference and an item's long-term characteristics by regularizing the local neighborhood representation for each node. \emph{Second}, a novel \textbf{global structure distillation} strategy is proposed to encode the global position for each user and item node. \emph{Third}, a general degree-aware \textbf{self-embedding distillation} component is applied to regularize the learned user and item embedding with a fixed quadratic constraint between the embedding learned on the history data and the embedding learned from new data. 

\subsection{Local Structure Distillation}
The key idea in GCNs~\cite{kipf2017, hamilton2017,ying2018,sun2020_mgcf} is to learn how to iteratively aggregate feature information from the local topology using neural networks. This aggregation step allows each node to benefit from rich contextual information and learn a more general node representation using information from the local neighborhood. Thus, it is crucial for an incrementally trained graph-based recommender system to preserve this local neighborhood representation. To achieve this objective, we proposed to distill the dot product value between the center node embedding and its neighborhood representation. 
For top-k recommender systems, the most representative information is the dot product between the user representation and item representation~\cite{deepfm, NGCF_wang19, ying2018, sun2020_mgcf}, which encodes the user's interest for this item. Taking a user node $u$ in the user-item bipartite graph as an example, the direct neighbour set of $u$ for time interval $t{-}1$ is $N_{u}^{t-1}$, which contains the items that this user interacted with during this time frame. 
When the new batch of data comes at time $t$, we want to transfer the dot product between its already trained user embedding $emb_u^{t-1}$ and the average embedding of its local neighborhood $c_{u,N_{u}^{t-1}}^{t-1}$ from the previous time frame to its incrementally trained student model. We achieve this by minimizing the difference of this dot product between the teacher model trained based on the data at $t{-}1$ and a student model trained using the most recent data at $t$.

The average local neighborhood embeddings $ c_{u,N_{u}^{t-1}}^{t-1}$ and $ c_{u,N_{u}^{t-1}}^t$ encode the general preferences for user $u$ at the previous time block and the current time block, respectively. By ensuring that $emb_u^{t-1}\cdot  c_{u,N_{u}^{t-1}}^{t-1}$ remains relatively close to $emb_u^t\cdot  c_{u,N_{u}^{t-1}}^t$, the model can explicitly preserve a user's historical preference. The same distillation scheme is applied on the item nodes. We define the local structure distillation loss term $\mathcal{L}_{local}$ as follows:
\begin{equation}
\small
\begin{split}
\mathcal{L}_{local} & =  \left( \frac{1}{|\mathcal{U}|} \sum_{u\in\mathcal{U}} (emb_u^{t-1} \cdot c_{u,N_{u}^{t-1}}^{t-1} - emb_u^{t} \cdot c_{u,N_{u}^{t-1}}^t)^2  \right.\\
& \left.+ \frac{1}{|\mathcal{I}|} \sum_{i \in \mathcal{I}} ( emb_i^{t-1} \cdot c_{i,N_{i}^{t-1}}^{t-1} - emb_i^{t} \cdot c_{i,N_{i}^{t-1}}^t)^2\right)\,,
\end{split}
\end{equation}
\begin{align}
\small
 c_{u,N_{u}^{t-1}}^t = \frac{1}{|\mathcal{N}_u^{t-1}|}\sum_{i' \in \mathcal{N}_u^{t-1}} emb_{i'}^{t}\\
 c_{i,N_{u}^{t-1}}^t = \frac{1}{|\mathcal{N}_i^{t-1}|}\sum_{u' \in \mathcal{N}_i^{t-1}} emb_{u'}^{t}\,.
\end{align}
Here $\mathcal{U}$ and $\mathcal{I}$ are the sets of user nodes and item nodes respectively. $|\mathcal{U}|$ is the total number of users and $|\mathcal{I}|$ the total number of items. $\mathcal{N}_u^t$ is the set of nodes in the neighbourhood of node $u$ at time $t$, and $|\mathcal{N}_u^t|$ is the degree of node $u$ at time $t$.  $c_{u,N_{u}^{t-1}}^{t-1}$ is the average embedding of the local neighbourhood of node $u$ at time $t{-}1$.

\subsection{Global Structure Distillation}

The local structure distillation scheme promotes the transfer of a teacher model's local topological information to a student model. However, it fails to capture each node's global position information, which is the relative position of the node with respect to all the other nodes. This is a general limitation of the majority of GNN models~\cite{you2019position,Srinivasan2020_equivalence_pos}. However, each node's global position encodes rich information for the recommendation setting. 

\begin{figure}[ht]
\small
    \centering
    \includegraphics[width=.95\linewidth]{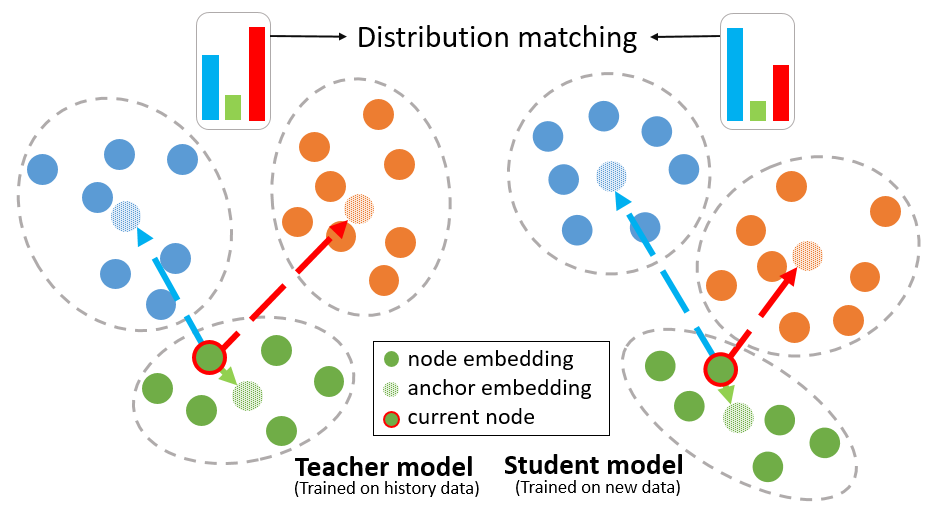}
    \caption{Global structure distillation demonstration}
\label{fig:global_structure_distiall}
\end{figure}

Consider a particular user node. The embedding distances between this user and all other users can encode the general user preference group of this user. The embedding distance between this user and all the items can encode which types of item this user likes. Thus, in the global structure distillation step, we aim to preserve the embedding information that encodes a node’s positional information with respect to all other nodes in the graph. To achieve this goal, we propose to use a set of anchor embedding on both the user side $\mathcal{A}_u$ and the item side $\mathcal{A}_i$ to encode the global structure information. These anchor embeddings are calculated using the average embeddings of clusters derived using $K$-means clustering of the user and item embeddings, respectively.  We obtain $2K$ clusters and each cluster represents a general user preference group or an item category. For each user node we then construct two probability distributions. The first captures the probability that a user belongs to a user preference group. The second represents the probability that an user favours a particular item category. Similar distributions are constructed for each item node. These probability distributions are constructed by considering the (normalized) embedding similarities to the anchor nodes. We achieve knowledge transfer by introducing a regularization term into the loss function that encourages matching of the distributions of the teacher with those of the student. We aim to minimize the sum of the Kullback-Leibler (KL) divergences between the teacher and student global structure distributions. Figure~\ref{fig:global_structure_distiall} illustrates the mechanism. For a user node, the global structure similarity between the teacher model and the student model can be computed as:


\begin{equation}
\small
S_{u,\mathcal{A}_u} = D_{KL}(GS_{u,\mathcal{A}_u^t}^{t}||GS_{u,\mathcal{A}_u^{t-1}}^{t-1}) = \displaystyle\sum_{k=1}^{K} GS_{u,\mathcal{A}_u^{t,k}}^t \log\left(\frac{GS_{u,\mathcal{A}_u^{t,k}}^t}{GS_{u,\mathcal{A}_u^{t-1,k}}^{t-1}}\right) ,
\end{equation}
where
\begin{equation}
GS_{u,\mathcal{A}_u^{t,k}}^t = \frac{e^{SIM(emb_u^t, \mathcal{A}_u^{t,k})/\tau}}{\displaystyle\sum_{k'=1}^{K} e^{SIM(emb_u^t, \mathcal{A}_u^{t,k'})/\tau}}, \hspace{0.2cm} SIM(z_i, z_j)= z_i^T z_j  \,.
\end{equation}

Here $\mathcal{A}_u^t$ and $\mathcal{A}_u^{t-1}$ are the sets of average embedding of each node cluster, which is clustered at time $t-1$, using $\{emb^t\}$ and $\{emb^{t-1}\}$ respectively. $GS_{u,\mathcal{A}_u^{t,k}}^t$ and $GS_{u,\mathcal{A}_u^{t-1,k}}^{t-1}$ are the $k$-th entries of the global structure distributions associated with the $K$ user anchor embedding contained in $\mathcal{A}_u^t$ and $\mathcal{A}_u^{t-1}$ for the student model and the teacher model, respectively.

To compute the final global structure distillation loss, we compute the average of the KL divergences between the distributions over all the nodes of the given graph:
\begin{align}
\small
  \mathcal{L}_{global} &= \left(\tfrac{1}{|\mathcal{N}_u|} \displaystyle\sum_{u\in\mathcal{U}} (S_{u,\mathcal{A}_u} + S_{u,\mathcal{A}_i}) +\right.\nonumber \\
  &\quad \quad \quad \quad \quad \quad \quad \quad \quad \left. \tfrac{1}{|\mathcal{N}_i|} \displaystyle\sum_{i\in\mathcal{I}} (S_{i,\mathcal{A}_u} + S_{i,\mathcal{A}_i})\right)\,.
\end{align}

\subsection{Self-embedding Distillation}
Finally, to preserve each user's and each item's own information (independent of the graph and neighbourhood), we directly distill the knowledge of each user's and item's embedding by adding mean squared error terms in the loss function. This ensures that each incrementally learned embedding does not move too far from its previous position. We control the distillation strength for each node using a weight factor $\eta$ which is proportional to the number of new records introduced for each node in the new data block. In other words, we enhance the distillation strength for nodes with richer historical records. The distillation loss term for self-embedding is:
\begin{equation}
\small
\begin{split}
\mathcal{L}_{self} = & \left( \frac{1}{|\mathcal{U}|} \sum_{u \in \mathcal{U}} \frac{\eta_u}{||\eta_{\boldsymbol{U}}||_2} ||emb_{u}^{t-1} - emb_{u}^t ||_2 \right. \\
& \left. + \frac{1}{|\mathcal{I}|} \sum_{i \in \mathcal{I}} \frac{\eta_i}{||\eta_{\boldsymbol{I}}||_2}  ||emb_{i}^{t-1} - emb_{i}^t ||_2\right) ,
\end{split}
\end{equation}
\begin{equation}
\small
\eta_u = \frac{|\mathcal{N}_u^{t-1}|}{|\mathcal{N}_u^t|}, \hspace{0.4cm} \eta_i = \frac{|\mathcal{N}_i^{t-1}|}{|\mathcal{N}_i^t|}\,,
\end{equation}
$\eta_u$ and $\eta_i$ are the coefficients that control the distillation strength of each node.

\subsection{Model Training}
We adapt our model to allow forward and backward propagation for mini-batches of triplet pairs $\{(u, i, j)\}$, where $i$ is a positive item (one the user has interacted with) and $j$ is a randomly sampled negative item. We optimize our model with respect to the widely-used Bayesian Personalized Ranking (BPR)~\cite{RendleFGS2009_bpr} loss:
\begin{equation}
\small
\begin{aligned}
\mathcal{L}_{BPR} = & \sum_{(u, i, j)\in\mathcal{O}} -\log\sigma( \mathbf{e}^*_u \cdot \mathbf{e}^*_i - \mathbf{e}^*_u \cdot \mathbf{e}^*_j)\ + \lambda_{l2}||\Theta||^{2}, 
\end{aligned}
\end{equation}
where $\mathcal{O}=\{(u,i,j)|(u,i)\in\mathcal{R^+}, (u,j)\in\mathcal{R^-})\}$ denotes the training batch, and $\Theta$ is the model parameter set. $\mathcal{R^+}$ indicates observed positive interactions and $\mathcal{R^-}$ indicates sampled unobserved negative interactions. 

The total loss combines the BPR loss, the self-embedding distillation loss, and the local and global structure preservation losses. 
\begin{equation}
\small
\mathcal{L} = \mathcal{L}_{BPR} +  \lambda_{self}\mathcal{L}_{self} + \lambda_{local}\mathcal{L}_{local} +  \lambda_{global}\mathcal{L}_{global}\,,
\end{equation}
where $\lambda_{self}$, $\lambda_{local}$, and $\lambda_{global}$ are hyperparameters that control the magnitude of the self-embedding, local structure distillation, and global structure distillation, respectively.

%% file: sections/experimental_results.tex
\section{Experiments}
\label{sec:results}

\subsection{Datasets}
To evaluate the effectiveness of GraphSAIL, we conduct experiments on two commonly used public datasets: \emph{Gowalla} and \emph{LastFM}, which vary in size, domain, time range, and sparsity. The statistics of the datasets are summarized in Table \ref{tab:dataset}. To simulate the real incremental learning scenario, we split both datasets by the absolute timestamp.

\begin{table}[t]
  \caption{Statistics of evaluation datasets.}
  \label{tab:dataset}
  \centering
  \resizebox{\columnwidth}{!}{
      \begin{tabular}{ccccccc}
        \toprule
                    & \#User    & \#Item     & \#Interaction  & \#Density  & \#Time span  & \#Inc block size \\
        \midrule
        \midrule
        Gowalla     & 29,858     & 40,998     & 1,027,464   & 0.0839\%  & 19 mths   & 61,647 \\
        \midrule
        LastFM      & 1,892      & 12,523     & 186,474     & 0.7870\%  & 6 yrs     & 18,647 \\
        \bottomrule
      \end{tabular}
  }
\end{table}

\textbf{Gowalla}\footnote{https://snap.stanford.edu/data/loc-gowalla.html}:
This is a real-world dataset collected from users' check-in records. We filter out duplicate interactions and user and item nodes with less than 10 records. We split the dataset such that the first 70\% forms a \emph{base block} and the remaining 30\% is divided evenly into 5 {\em incremental blocks}. 

\textbf{LastFM}\footnote{https://grouplens.org/datasets/hetrec-2011/}:
This is a real-world dataset collected from users' tagging of artists' records and commonly used for recommending artists to users. Due to the small size of the dataset, we do not filter out any interaction or node. To ensure sufficient training data for each incremental block, we split the dataset as follows: the first 60\%  as the \emph{base block} and the remaining 40\% into 4 \emph{incremental blocks}. 

\subsection{Experiment Setting}

\subsubsection{Evaluation metrics}

\begin{figure}[ht]
    \centering
    \includegraphics[width=\linewidth]{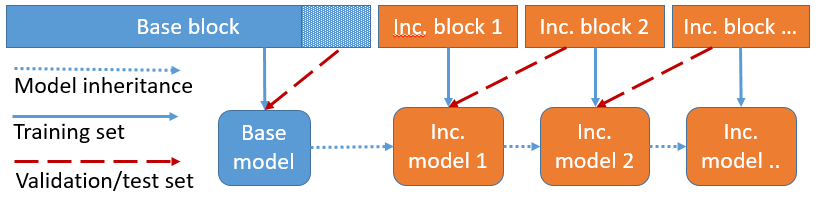}
    \caption{Incremental experiment setting. For the \emph{base block}, solid area is for training and shaded area is a random subset for test and validation. For \emph{incremental blocks}, block $t$ is used for incremental training and block $t+1$ is used for test and validation.}
    
\label{fig:data_split}
\end{figure}

To evaluate the quality of the recommendation results, we evaluate our method and baselines using \emph{Recall@20}, which is a commonly used evaluation metric in top-k recommendation~\cite{NGCF_wang19,sun2020_mgcf}. Users' preference scores are predicted over all items except the existing items in the training set. \emph{Recall@20} then measures the fraction of the total amount of positive items that were discovered by the top 20 predicted items of each user.

To evaluate the effectiveness of the incremental training and also simulate the real scenario, we use \emph{incremental block} $t$ for training and \emph{incremental block} $t+1$ for test and validation. We then take the average of the test scores over all consecutive incremental blocks as the final evaluation metric for incremental methods. More specifically, as shown in Figure \ref{fig:data_split}, for the \emph{base block}, we randomly split 80\% of interactions for training and 10\% each for validation and test. For incremental block 1(Inc. b1), we use the entire data block for training and randomly split Inc. b2 to two halves for validation and test. Note we only use the first incremental block to tune hyperparameters and keep them fixed for the following incremental blocks. More details about hyperparameter selection are described in Section \ref{sec:parameter}. 

\subsubsection{Recommender base models}

To demonstrate the effectiveness of GraphSAIL on recommender models, we adapt the following two GNN-based recommender models to the incremental learning setting. The two base models vary in architecture and learn different graph information.

\textbf{PinSage~\cite{ying2018}}:
PinSage is a recommender model that uses GraphSAGE~\cite{hamilton2017} on the item-item graph. Inspired by~\cite{sun2020_mgcf} , we adapt it to the user-item bipartite graph by using two PinSage networks for user nodes and item nodes separately. 

\textbf{MGCCF~\cite{sun2020_mgcf}}:
MGCCF is a recent state-of-the-art top-k recommender model that leverages GCNs to learn user-item, user-user, and item-item graph information simultaneously.

\subsubsection{Baseline Algorithms}

To demonstrate that our proposed framework can preserve graph structure information and alleviate catastrophic forgetting, we compare our method with the following baseline algorithms.

\textbf{Fine-tune (FT)}: Fine-tune is a naive baseline that uses solely the new data to fine-tune the previously trained model.

\textbf{Embedding distillation (Emb\_d)}: Embedding distillation is an unweighted version of self information distillation. It distills node embedding information by applying a mean squared error regularization term to all node embedding in the loss function.

\textbf{LSP\_s~\cite{yang_distillatingGCN}}: LSP is a recent work that applies knowledge distillation to GCNs by preserving the local structure of each node; it achieves state-of-the-art performance on point cloud data. We modify the original LSP to a scalable version, which can be applied to large graphs, by randomly selecting 10 neighbours for each node for its local structure preservation.

\textbf{Full-batch (Fu\_batch)}: Full batch training uses all the historical data including the base block to train the model from scratch at each incremental step. It is used as a reference to show the full-data performance and training time at each incremental step.

\begin{table}[ht]
  \caption{Hyperparameters used during incremental learning. $\lambda_{self}$, $\lambda_{local}$, $\lambda_{global}$, K, $\tau$ are the hyperparameters used by GraphSAIL. $\lambda_{Emb\_d}$ and $\lambda_{LSP\_s}$ are the distillation coefficients used by Emb\_d and LSP\_s, respectively.}
  \label{tab:hyperparameters}
  \centering
  \resizebox{\columnwidth}{!}{
      \begin{tabular}{c|c|ccccccc}
        \toprule
        \multicolumn{2}{c}{} & $\lambda_{self}$ & $\lambda_{local}$ & $\lambda_{global}$ & K & $\tau$ & $\lambda_{Emb\_d}$ & $\lambda_{LSP\_s}$ \\
        \midrule
        \midrule
        \multirow{2}{*}{LastFM}    & PinSage      & 100    & 1e5   & 1e6   & 10     & 0.1     & 10    & 1e9  \\
                                   & MGCCF      & 100    & 1e7   & 1e7   & 5      & 0.1     & 10    & 1e10  \\
                                   
        \midrule
        \multirow{2}{*}{Gowalla}    & PinSage     & 10     & 1e6   & 1e3   & 20     & 1       & 1     & 1e6  \\
                                    & MGCCF     & 100    & 1e4   & 1e4   & 10     & 1       & 10    & 1e4  \\
      \bottomrule
      \end{tabular}
    }
\end{table}

\begin{table*}[ht]
  \caption{The overall performance comparison. Results are averages of 10 repeated experiments. \%imprv is the relative improvement with respect to the fine-tune result. Training time is the average training time for one incremental block. $\star$ denotes $p<0.05$ when performing the two-tailed pairwise t-test on GraphSAIL with the best baseline.}
  \label{tab:result}
  \scriptsize{
  \begin{tabular}{c|c|cccccccc}
    \toprule
    \multicolumn{2}{c}{} & Algo. & Inc. b1 & Inc. b2 & Inc. b3 & Inc. b4 & Avg. Recall@20 & \%imprv & Training time \tablefootnote{Training time refers to the time needed to train the model until convergence (excluding the test time on validate data). To compare the efficiency fairly, we conduct experiments of different models on the same machine (with a single NVIDIA Tesla V100 GPU).} \\
    \midrule
    \midrule
    \multirow{10}{*}{LastFM} & \multirow{5}{*}{PinSage}  & FT        & 0.0324 & 0.0516 & 0.0463 & - & 0.0434 $\pm$ 0.0023 &    - &  8 s \\
                             &                          & Emb\_d    & 0.0348 & 0.0556 & 0.0488 & - & 0.0464 $\pm$ 0.0018 & 6.91\%  &  3 s \\
                             &                          & LSP\_s    & 0.0360 & 0.0517 & 0.0446 & - & 0.0441 $\pm$ 0.0021 & 1.61\%  &  8 s \\
                             &                          & GraphSAIL$^{\star}$ & \textbf{0.0372} & \textbf{0.0563} & \textbf{0.0498} & - & \textbf{0.0478} $\pm$ 0.0018 & 10.14\% & 5 s\\        
                                                        \cline{4-10}
                              &                         & Fu\_batch \tablefootnote{Batch training results are of 1 experiment, not 10 repetitive experiments. }  & 0.0345 & 0.0509 & 0.0502 & - & 0.0452              & - &  2 min \\
                                 \cline{2-10}
                            &   \multirow{5}{*}{MGCCF}    & FT         & 0.0380 & 0.0481 & 0.0513 & - & 0.0458 $\pm$ 0.0016 & - &  7 s \\
                            &                             & Emb\_d     & 0.0410 & 0.0483 & \textbf{0.0539} & - & 0.0477 $\pm$ 0.0018 & 4.15\%  & 9 s \\
                            &                             & LSP\_s     & 0.0402 & 0.0504 & 0.0462 & - & 0.0456 $\pm$ 0.0017 & -0.44\% & 5 s \\
                            &                             & GraphSAIL$^{\star}$  & \textbf{0.0422} & \textbf{0.0532} & 0.0518 & - & \textbf{0.0491} $\pm$ 0.0014 & 7.21\%  & 7 s \\        
                                                        \cline{4-10}
                            &                            & Fu\_batch \footnotemark[4]  & 0.0483 & 0.0554 & 0.0566 & - & 0.0534              & - & 4 min\\                                
                            
    \midrule
    \multirow{10}{*}{Gowalla} & \multirow{5}{*}{PinSage} & FT        & 0.1001 & 0.1032 & 0.1056 & 0.1276 & 0.1091 $\pm$ 0.0012 &  - & 8 s \\
                             &                          & Emb\_d    & 0.0984 & 0.1027 & 0.1051 & 0.1206 & 0.1067 $\pm$ 0.0013 & -2.20\% & 6 s \\
                             &                          & LSP\_s    & \textbf{0.0999} & 0.1050 & 0.1075 & \textbf{0.1292} & \textbf{0.1104} $\pm$ 0.0017 & 1.19\%  & 4 s \\
                             &                          & GraphSAIL & 0.0998 & \textbf{0.1056} & 0.1075 & 0.1262 & 0.1098 $\pm$ 0.0019 & 0.64\%  & 27 s \\              
                                                        \cline{4-10}
                              &                         & Fu\_batch \footnotemark[4]  & 0.1126 & 0.1195 & 0.1273 & 0.1651 & 0.1311              & - & 62 min \\  
                                \cline{2-10}
                            &   \multirow{5}{*}{MGCCF}  & FT       & 0.1018 & 0.1058 & 0.1120 & 0.1400 & 0.1149 $\pm$ 0.0009 & - &  22 s\\
                            &                           & Emb\_d   & 0.1002 & 0.1031 & 0.1077 & 0.1300 & 0.1103 $\pm$ 0.0008 & -4.00\% & 22 s\\
                            &                           & LSP\_s   & 0.1048 & 0.1069 & 0.1148 & 0.1415 & 0.1170 $\pm$ 0.0012 & 1.83\%  & 25 s\\
                            &                           & GraphSAIL$^{\star}$  & \textbf{0.1065} & \textbf{0.1086} & \textbf{0.1169} & \textbf{0.1456} & \textbf{0.1194} $\pm$ 0.0009 & 3.92\%  & 42 s\\ 
                                                        \cline{4-10}
                            &                           & Fu\_batch \footnotemark[4]   & 0.1142 & 0.1159 & 0.1291 & 0.1694 & 0.1254              & - & 148 min \\
  \bottomrule
\end{tabular}}
\end{table*}

\subsubsection{Parameter Settings}
\label{sec:parameter}

All the base models and baseline algorithms are implemented with TensorFlow. The embedding dimension of all nodes is set to 128 and the Adam optimizer with the Xavier initialization is used for all experiments. For both base models, we adopt the hyperparameters used in the original MGCCF implementation and keep them fixed throughout all experiments, except for the learning rate. 
For PinSage, 10 random neighbours are selected for the first layer aggregation. For MGCCF, 10 and 5 random neighbours are selected for the first and second layer of aggregation respectively. We also use a negative sampling strategy to randomly sample 10 unobserved interactions to be paired with each positive interaction for both base models for training.

For the base blocks, we set the learning rate to 1e-3 and we set early stop to 50, i.e., we stop training when the \emph{recall@20} stops increasing on the validation set for 50 epochs. For the incremental blocks, we reduce the learning rate to 5e-4 and we set early stop at 2 and train each incremental block up to 15 and 6 epochs for Gowalla and LastFM respectively. To simulate the real incremental scenario, other hyperparameters are optimized using grid search on the validation part of Inc. b1 and remain fixed for the following incremental blocks. The hyperparameters are listed in Table \ref{tab:hyperparameters}.

\subsection{Comparison with Baselines}

Table \ref{tab:result} reports the performance comparison of GraphSAIL and other baselines. We make the following observations:

\begin{enumerate}[label=(\arabic{enumi}),ref=\arabic{enumi}, wide=0pt, topsep=4pt]
    \item GraphSAIL outperforms the fine-tune baseline for all base models and datasets. More precisely, taking the average \emph{Recall@20} over all incremental blocks, on the LastFM dataset it outperforms the fine-tune method by 10.14\% and 7.21\% with PinSage and MGCCF, respectively; on the Gowalla dataset, it outperforms the fine-tune method by 0.64\% and 3.92\% with PinSage and MGCCF, respectively.
    
    \item GraphSAIL yields the best average \emph{Recall@20} over all baseline methods except when using PinSage as the base model on the Gowalla dataset. However, in this experiment the difference between all baselines and the fine-tune method are within one standard deviation of the 10 repeated experiments. We argue that this is because the one layer PinSage model fails to quickly learn from the incremental block and thus there is only minimal catastrophic forgetting in this case. As a result, distilling the previously learned knowledge does not improve the performance. This can be validated by comparing the performance of the two base models on the Gowalla dataset. At the last incremental block, the \emph{Recall@20} of using GraphSAIL on MGCCF outperforms the \emph{Recall@20} of using GraphSAIL on PinSage by 15.37\%.
    
    \item By looking at the result of Inc. b1, which is the only block where hyperparameters are optimized, we see that GraphSAIL and LSP\_s always outperform the fine-tune method, except for the case discussed above. This demonstrates that catastrophic forgetting can be alleviated by distilling previously learned knowledge by considering the graph structure. GraphSAIL performs the best because, in contrast to LSP\_s, it also captures the global graph structure.
    
    \item The hyperparameters are only optimized for Inc. b1, and we cannot guarantee that users' preferences drift (or items' properties drift) at a constant rate for the following blocks. We hope to have a robust method that can alleviate catastrophic forgetting despite these uncertainties. By looking at the average \emph{Recall@20} over all incremental blocks, we see that Emb\_d only works for LastFM and LSP\_s only works for Gowalla, whereas GraphSAIL constantly yields good performance except for the case discussed above.
    
    \item On the Gowalla dataset, incremental learning can reduce the training time by over 2 orders of magnitude compared to full-batch training. For a large scale dataset, updating the model incrementally makes it possible to greatly increase the model update frequency.
\end{enumerate}

\subsection{Ablation Studies}

To further examine the effectiveness of each component of GraphSAIL, we additionally conduct experiments on LastFM by adding different combinations of components to the fine-tune baseline. Table~\ref{tab:ablation} shows the experimental results for Inc. b1 for which the hyperparameters are optimized. First, we observe that the self-information distillation component makes a more substantial contribution for PinSage than for MGCCF. We conjecture that this occurs because the one layer PinSage model is a simple GCN model which has a lower capability to capture graph structure. Thus directly distilling the node embedding, without considering graph structure, plays a more important role. Second, we observe that for the MGCCF model, each component of GraphSAIL makes a contribution to the performance improvement. Last, although the local structure distillation component has a negative impact to PinSage, the performance of using all three components (\emph{FT+self+local+global}) is better than using any subset (e.g. \emph{FT+self+global}) for both models. The local structure distillation alone performs poorly here because hyperparameters are tuned for the combination of all components.
Moreover, we see that preserving the global structure boosts the performance significantly for both models.

\begin{table}[ht]
  \caption{Ablation analysis of applying different components on LastFM for one incremental block. Results are averages of 10 repeated experiments. \%imprv is the relative improvement with respect to the fine-tune result.}
  \label{tab:ablation}
  \centering
  \scriptsize{
  
      \begin{tabular}{c|cc|cc}
        \toprule
         & \multicolumn{2}{c|}{PinSage} & \multicolumn{2}{c}{MGCCF} \\
         \cline{2-5}
        Algo. & Inc. b1 & \%imprv & Inc. b1 & \%imprv \\
        \midrule
        \midrule
        \multicolumn{1}{l|}{FT }                    & 0.0324 &    -     & 0.0380 &-         \\
        \multicolumn{1}{l|}{FT+self }               & 0.0337 & 4.01 \%  & 0.0386 & 1.58 \%  \\
        \multicolumn{1}{l|}{FT+local }              & 0.0310 & -4.32 \%  & 0.0383 & 0.79 \%  \\
        \multicolumn{1}{l|}{FT+global }             & 0.0356 & 9.88 \%  & 0.0395 & 3.95 \%  \\
        \multicolumn{1}{l|}{FT+self+local }         & 0.0337 & 4.01 \%  & 0.0400 & 5.26\%  \\
        \multicolumn{1}{l|}{FT+self+global }        & 0.0355 & 9.57\%  & 0.0398 & 4.74 \%  \\
        \multicolumn{1}{l|}{FT+local+global }       & 0.0358 & 10.49 \%  & 0.0404 & 6.32 \%  \\
        \multicolumn{1}{l|}{FT+self+local+global}   & \textbf{0.0372} & \textbf{14.81 \%}  & \textbf{0.0422} & \textbf{11.05 \%} \\
      \bottomrule
      \end{tabular}}
\end{table}

\subsection{Comparison between Graph-based and Non-Graph Models}

To better understand the advantages of using graph-based model to do incremental learning for recommender systems, we conduct the comparison experiments of using the basic fine-tuning method of 10 epochs to update a graph-based model and a non-graph model, namely MGCCF and BPRMF~\cite{RendleFGS2009_bpr}, on LastFM. In Figure~\ref{fig:BPRvsMGCCF}, we observe that when directly predicting for the future interactions, i.e., the beginning points of the green and the yellow curves, the embedding learned by MGCCF has a better performance. Moreover, during the fine-tuning phase, the graph-based model quickly converges to the peak point, whereas the non-graph model cannot learn from the new data effectively. Last, there is a clear sign of catastrophic forgetting when fine-tuning using MGCCF but not for BPRMF. This also verifies that a graph-based model learn more efficiently than a non-graph method when given a small amount of new data. An explanation is that updating a node embedding using a graph-based model leads to implicitly updating its neighbours. Therefore, a node embedding could be correctly (usefully) updated even if the node does not exist in the new data when the graph information is considered. Overall, we argue that applying incremental learning on graph-based models has a greater potential than applying it on conventional recommendation models.

    


%% file: sections/application.tex
\section{Application: Candidate Selection for App Recommendation}
\begin{table*}   
\caption{Performance Comparison (Logloss) in Industrial Dataset. $\star$ denotes $p<0.05$ when performing the two-tailed pairwise t-test on GraphSAIL with the best baseline.} 
\label{tab:industrial}
\scriptsize{ \begin{tabular}{ccccccccc}
    \toprule
    & Inc. day 1 & Inc. day 2 & Inc. day 3 & Inc. day 4   & Inc. day 5 & Inc. day 6 & \% Avg. Improv & Training time \\
    \midrule
batch mode  &  0.3321    & 0.3057 &0.4307 & 0.3677 & 0.3818 & 0.3781 & - & 912.94 s \\
 FT    &  0.3232& 0.2927 &0.4135  &0.3625  & 0.3651 & 0.3613 & 3.53\% & 30.68 s \\
 Emb\_d         & 0.3146 & 0.2836 & 0.4032 & 0.3582 & 0.3645 & 0.3622 & 5.03\% & 30.96 s\\
 LSP\_s & 0.3164 & 0.2872 & 0.4021 & 0.3618 & 0.3635 & 0.3604 & 4.75\% & 31.27 s  \\  
 GraphSAIL$^{\star}$ & \textbf{0.3127} & \textbf{0.2824} & \textbf{0.3984} & \textbf{0.3536} & \textbf{0.3612} & \textbf{0.3597} & 5.84\% & 31.94 s \\
                
\bottomrule
\end{tabular}}
\end{table*}
To validate the effectiveness of GraphSAIL, we deploy it in the recommender system of a mainstream App Store. As shown in Figure~\ref{fig:online_pipeline}, the recommender system consists of three modules: candidate selection, matching, and ranking. In our App Store, there exist tens or hundreds of thousands of apps. If all the apps are involved for recommendation, the latency is unacceptable in practice, because it takes time to predict the matching score between a user and all the apps. Therefore, the candidate selection module is needed to select several hundred apps from the universal set for each user. Then, the matching module predicts the matching score between a user and the selected apps representing the user's preference on them. Lastly, the ranking module produces the ranking list based on the matching scores and other principles (diversity and business rules). 


\begin{figure}[ht]
    \centering
    \includegraphics[width=0.65\linewidth]{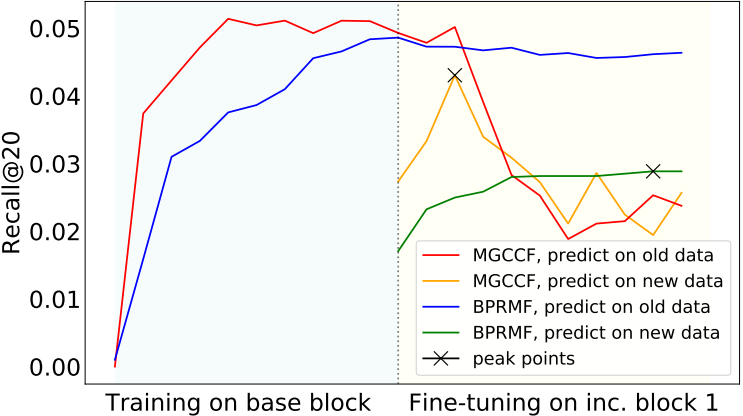}
    \caption{Fine-tuning analysis of MGCCF and BPRFM on LastFM. The left part of the figure shows the validation curves of MGCCF and BPRMF when training on base blocks. In the right part of the figure, both models are fine-tuned using solely the incremental data. The two continuous curves show the Recall@20 on the validation set of the base block. The two new curves show the Recall@20 on the validation set of the incremental block.}
    
\label{fig:BPRvsMGCCF}
\end{figure}

\subsection{Dataset and Compared Models} 

\begin{figure}[!ht]
    \centering
    \includegraphics[width=0.65\linewidth]{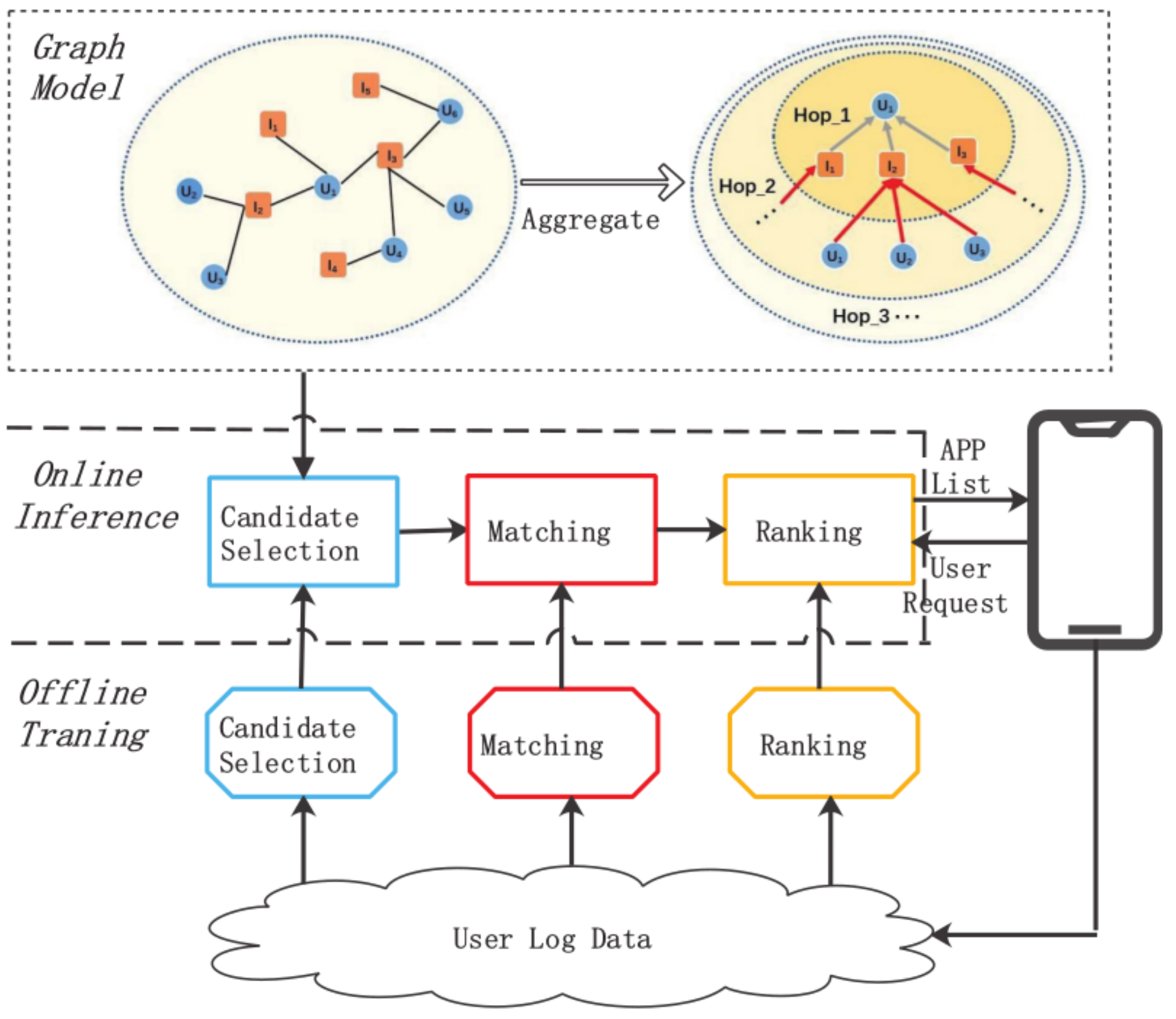}
    \caption{Pipeline of the APP recommendation system.}
\label{fig:online_pipeline}
\end{figure}

We sample and collect 22 consecutive days of user-app download records from the game recommendation scenario. A user-item bipartite graph is constructed based on the download records. Two types of nodes (user nodes and item nodes) exist and an edge between a user node and an item node is included if there is an associated download record. App features (e.g., identification and category) and user features (e.g., click/browse/search history) are used as node features in the graph. As can be observed from Figure~\ref{fig:online_pipeline}, a GCN model (i.e., a simplied version of MGCCF) has been deployed in the candidate selection module. MGCCF learns an embedding for each user as well as each item. Then, candidates for each user are generated by finding the most similar items by computing the similarity between user and item embedding.


The model employed in the candidate selection module of our recommender system is trained on 14 days' data, which is referred to as ``batch mode''. Based on the collected 22 days' data, we test the performance of ``batch mode'' in a sliding window manner. Specifically, days $[T,T+13]$ of data are used for training, day $T+14$ of data is used for validation, and day $T+15$ of data is for testing, where $T\in[1,6]$.
On the other hand, the model trained by different incremental learning strategies starts with a warm-start model (which is trained on days $[0,13]$ of data), instead of from scratch. The model is then incrementally updated with training data day $T$, validated on data $T+1$, and tested on data $T+2$, where $T\in[14,19]$.

\subsection{Performance Comparison}
The performance comparison on the industrial dataset is presented in Table \ref{tab:industrial}. The effectiveness of a model is measured by Logloss, while the efficiency is measured by the model's training time in seconds. 
As can be observed, our model achieves consistently better results compared to baseline models on all incremental days.
Specifically, GraphSAIL outperforms batch mode by 5.84\% in terms of Logloss. For efficiency concern, it takes just 3.5\% of the training time of batch mode approach. Moreover, GraphSAIL outperforms the other incremental learning baselines which is consistent with the observations in public datasets.
It demonstrates the advantage of our model for incrementally training a GCN model on a large-scale industrial scenario with rich side information.

%% file: sections/conclusion.tex
\section{Conclusion}

In this work, we propose a novel graph structure aware incremental training paradigm GraphSAIL to update GNN-based recommender models. GraphSAIL counters catastrophic forgetting by explicitly preserving the node self-information, local structure information and global structure information of GNN models. We conducted experiments with two different GNN recommender models on two public datasets to demonstrate the effectiveness of GraphSAIL. We further deploy GraphSAIL on a real industrial dataset and achieve better performance than all baselines. Moreover, we demonstrate the potential of incrementally updating graph-based recommender models. This work serves as an early attempt to effectively apply incremental learning on GNN-based recommender systems.